\documentclass[aps,twocolumn,showpacs]{revtex4}
\usepackage{amssymb}

\usepackage{amsmath}
\usepackage{graphicx}

\begin{document}

\title{Spin squeezing criterion with local unitary invariance}
\author{A. R. Usha Devi$^{1,2}$, Xiaoguang Wang$^{1}$, and B. C. Sanders$%
^{1} $ }
\affiliation{1. Department of Physics and Australian Centre for Quantum Computer
Technology, \\
Macquarie University, Sydney, New South Wales 2109, Australia.}
\affiliation{2. Department of Physics, Bangalore University, Jnanabharathi Campus,
Bangalore 560 056, India.}
\date{\today}

\begin{abstract}
We propose a spin squeezing criterion for arbitrary multi-qubit states that
is invariant under local unitary operations. We find that, for arbitrary
pure two-qubit states, spin squeezing is equivalent to entanglement, and
multi-qubit states are entangled if this new spin squeezing parameter is
less than 1.
\end{abstract}

\maketitle

\section{Introduction }

The non-classical nature of quantum entanglement is a key ingredient in the
rapidly developing science of quantum information~\cite{Nie00,Bou00}. One of
the important tasks involved is quantifying quantum entanglement, which is
essential in assessing the performance of a quantum system in several
applications such as quantum teleportation~\cite{Ben93}, quantum
cryptography~\cite{Eke91, Cir97, Fuc97}, quantum computation ~\cite%
{Bra99,Eke98,Ben92,Eis00} and quantum communication~\cite{Eis00, Cir99}.
There has been an ongoing effort devoted to characterizing quantum
entanglement, especially the entanglement of multi-particle states shared by
several distant parties~\cite{Thapliyal99, Kem99, Ben99, Dur99, DiV99,
Smo00, Dur00, Shor00, Wong01}. In view of the recent interest in creating
and manipulating correlated collective atomic states~\cite{Win92, Kuz97,
Kuz98, Pol99, Hald99, Kuz00, Duan00, Koz00, Sor01a}, an experimentally
relevant characterization of entanglement in terms of \emph{\ spin squeezing}%
~\cite{Kit93,Win92,
Lukin,Vernac,Youli,kasevich,uffe,Ber02,Dominic,Usha,Gasenzer,Stockton,Add1,Add2,WangSpin}
has been highlighted ~\cite{Sor01a,Sor01b}. Most studies are concentrated on
collective symmetric multi-particle states where individual particles are
assumed to be inaccessible. Our aim is to develop a spin squeezing criterion
for a multi-qubit system that is invariant under local unitary
transformations on individual qubits. The motivation for this work is to
explore the fundamental connection between spin squeezing and entanglement,
recognizing that entanglement is invariant under local unitary
transformation.

Let us first review spin squeezing criteria. Several definitions of spin
squeezed states have been proposed in the literature ~\cite{Sor01a, Kit93,
Wod85, Aga90, Win94}. Kitagawa and Ueda~\cite{Kit93} pointed out that a
definition of spin squeezing, based only on the uncertainty relation~\cite%
{Wod85}, does not reveal quantum correlations among the elementary spins.
They first identified a mean spin direction 
\begin{equation}
\hat{n}_0=\frac{\langle\vec{J}\rangle} {|\langle\vec{J}\rangle|}, \quad {%
|\langle\vec{J}\rangle|}=\sqrt{\langle\vec{J}\rangle\cdot\langle\vec{J}%
\rangle},
\end{equation}
where the collective spin operator $\vec{J}$ for an $N$-qubit system is
defined by $\vec{J}\equiv\frac{1}{2}\displaystyle\sum_{i=1}^N \vec{\sigma}%
_{i}$ with $\vec{\sigma}_{i}=(\sigma_x,\sigma_y,\sigma_z)^T$ the Pauli
vector of the $i^{\mathrm{th}}$ qubit. Associating a mutually orthonormal
set $(\hat{n}_{\perp}, \hat{n}_{\vdash},\hat{n}_0)$, we have another set of
collective operators 
\begin{equation}
J_\mu=\vec{J}\cdot\hat{n}_\mu, \; \mu \in \{\perp,\vdash,0\} ,
\end{equation}
which satisfy the angular momentum commutation relations. Kitagawa and Ueda~%
\cite{Kit93} proposed that a multi-qubit state can be regarded as spin
squeezed if the minimum of $\Delta J_{\theta}$ of a spin component normal to
the mean spin direction is smaller than the standard quantum limit $\sqrt{N}%
/2$ of the spin coherent state, where 
\begin{equation}
J_{\theta}=J_\perp \cos\theta+ J_\vdash \sin\theta.
\end{equation}
A parameter incorporating this feature may be defined by 
\begin{equation}  \label{eq:Kitagawa}
\xi_{1}\equiv\frac{2\ (\Delta J_{\theta})_{\text{min}}}{\sqrt{N}},
\end{equation}
where the minimization is over $\theta$. A spin squeezed state satisfies $%
\xi_{1}<1$.

In the context of Ramsey spectroscopy on a sample of $N$ two-level atoms,
Wineland \textit{et al.}~\cite{Win94} showed that the frequency resolution
depends on the parameter 
\begin{equation}  \label{eq:Wineland}
\xi_2=\frac{\sqrt{N}\ (\Delta J_{\theta})_{\text{min}}}{\langle J_0 \rangle}
=\frac{2\, N\, \xi_{1}}{\langle J_0\rangle},
\end{equation}
and spin squeezing manifested by $\xi_2<1$ leads to reduction in the
frequency noise. This identification opened up possible applications of spin
squeezed states to high precision atomic clocks~\cite{Win94} and atomic
interferometers~\cite{Yur86}. Note that spin squeezing established through $%
\xi_{1}<1$ is a necessary condition for $\xi_2<1$, but it is not a
sufficient condition since $\langle J_0\rangle\leq 2N$.

Spin squeezing, in the original sense, is defined for multi-qubit states
belonging to the maximum multiplicity subspace of the collective angular
momentum operator $\vec{J}$. These states exhibit symmetry under the
interchange of particles. The concept of spin squeezing is therefore
restricted to symmetric multi-particle systems that are accessible to
collective operations alone. We explore the possibility of extending the
concept of spin squeezing to multi-qubit systems, where individual qubits
are accessible. This requires a criterion of spin squeezing that exhibits
invariance under local unitary operations on the qubits. In contrast, Eqs.~(%
\ref{eq:Kitagawa}) and (\ref{eq:Wineland}) are not invariant under arbitrary
local unitary transformations on the qubits. In order to see this, consider
a state of two qubits given by 
\begin{equation}
|\Psi _{12}\rangle =\cos \phi |01\rangle +\sin \phi |10\rangle ,
\label{eq:psi}
\end{equation}%
in which the expectation value of the collective angular momentum $\langle 
\vec{J}\rangle =0$, and therefore one cannot properly define spin squeezing
for this state. However, under a local unitary operation $U_{1}\otimes U_{2}=%
\openone\otimes \sigma _{x}$, the state $|\Psi _{12}\rangle $ transforms to 
\begin{equation}
|\Psi _{12}^{\prime }\rangle =\cos \phi |00\rangle +\sin \phi |11\rangle ,
\label{eq:psi'}
\end{equation}%
where $\openone$ is a $2\times 2$ identity matrix. One can readily verify
that the state $|\Psi _{12}^{\prime }\rangle $ is spin squeezed with the
squeezing parameters of Eqs.~(\ref{eq:Kitagawa}) and (\ref{eq:Wineland})
given by 
\begin{equation}
\xi _{1}=\sqrt{1-|\sin 2\phi |}\leq 1,\quad \xi _{2}=\frac{1}{\sqrt{1+|\sin
2\phi |}}\leq 1.
\end{equation}%
Therefore, the spin squeezing by either criterion is modified by a local
unitary transformation.

A simpler example is a product state $|\psi_1\rangle|\psi_2\rangle$, which
is in general non-symmetric. For the non-symmetric case, we find that the
squeezing parameter $\xi_1$ can be less than 1. This fact implies that the
squeezing parameter $\xi_1$ works well for symmetric states; however, for
non-symmetric cases, one cannot distinguish a correlated state from an
uncorrelated one.

In this paper, we provide a local unitary invariant version of spin
squeezing criteria, which reveals quantum correlations for arbitrary
multi-qubit states. It is well-known that quantum entanglement is locally
unitary invariant. The local unitary invariant property of the spin
squeezing criteria and quantum entanglement suggests that they may exhibit
closer relations comparing with relations between the original spin
squeezing criteria and entanglement.

The paper is organized as follows: In Sec.~II, we introduce two spin
squeezing criteria, which are shown to be locally unitary invariant. In
Sec.~III, we provide relations between spin squeezing and quantum
entanglement, and find that (i)~for arbitrary two-qubit pure states, spin
squeezing implies entanglement, and vice versa, and (ii)~for an arbitrary
multi-qubit state, if the local invariant version of squeezing parameter ${%
\xi }_{2}$ is less than 1, then the state is entangled. The conclusion is
given in Sec.~IV.

\section{Local unitary invariant spin squeezing criteria}

Now we introduce local unitary invariant spin squeezing criteria for $N$
qubits. We denote unit vectors along the mean orientations (mean spin
directions) of the qubit $i$ by 
\begin{equation}
\hat{n}_{i0}=\frac{\langle \vec{\sigma}_{i}\rangle} {|\langle\vec{\sigma}%
_{i}\rangle|}, \quad {|\langle\vec{\sigma}_{i}\rangle|}={(\langle\vec{\sigma}%
_{i}\rangle\cdot \langle\vec{\sigma}_{i} \rangle)^{\frac{1}{2}}}.
\label{angular0}
\end{equation}
Associating a mutually orthogonal set $(\hat{n}_{i\perp}, \hat{n}_{i\vdash},%
\hat{n}_{i0})$ of unit vectors with each qubit, we may define the collective
operators 
\begin{equation}
{}_{\perp}=\frac{1}{2}\sum_{i=1}^N \vec{\sigma}_i\cdot\hat{n}_{i\perp},\;
{}_{\vdash }=\frac{1}{2}\sum_{i=1}^N \vec{\sigma}_i\cdot\hat{n}_{i\vdash},\;
{}_0=\frac{1}{2}\sum_{i=1}^N \vec{\sigma}_i\cdot\hat{n}_{i0},
\label{angular}
\end{equation}
which satisfy the usual angular momentum commutation relations. For
instance, 
\begin{equation}
[{}_{\perp}, {}_{\vdash}]= \frac{\mathrm{i}}{2} \sum_{i=1}^N \vec{\sigma}%
_i\cdot \hat{n}_{i0}=\mathrm{i}\, {}_0,
\end{equation}
and this leads to the uncertainty relations 
\begin{equation}  \label{eq:Uncertainty}
(\Delta {}_{\perp}) (\Delta {}_{\vdash}) \geq \frac{1}{2}\,\langle
{}_0\rangle .
\end{equation}

Analogous to the definitions of $\xi_k,~k\in \{1,2\}$, we define the two
spin squeezing parameters $\tilde{\xi}_k$, 
\begin{equation}  \label{eq:NSpSqGen}
\tilde{\xi}_{1}=\frac{2 (\Delta {}_{\{\theta_i\}}) _{\mathrm{min}}} {\sqrt{N}%
}, \quad \tilde{\xi}_{2}=\frac{\sqrt{N}\, (\Delta{}_{\{\theta_i\}})_{\mathrm{%
min}}} {\langle {}_0\rangle},
\end{equation}
where 
\begin{equation}
{}_{\{\theta_i\}}=\frac{1}{2}\sum_{i=1}^N \vec{\sigma}_i\cdot \hat{n}%
_{\theta_i}, \; \hat{n}_{\theta_i}=\hat{n}_{i\perp} \cos\theta_i+\hat{n}%
_{i\vdash} \sin\theta_i,
\end{equation}
and the minimization is over all $\theta_i,~i\in\{1,2,\ldots,N\}$. They
satisfy local unitary invariance as we show below. Therefore, we have given
locally invariant versions of the original spin squeezing criteria. A system
of $N$ qubits are regarded as spin squeezed if $\tilde{\xi}_{k}<1$ for $%
k\in\{1,2\}$.

Now we show the key property of the spin squeezing parameter $\tilde{\xi}_k$
given by\newline
\textbf{Proposition 1}: \textit{The spin squeezing parameters $\tilde{\xi}%
_{k}$ are invariant under the local unitary operator 
\begin{equation}
U=U_1\otimes U_2\otimes\cdots\otimes U_N.
\end{equation}
}\newline
\textit{Proof}: From the expressions of $\tilde{\xi}_k$ given by Eq.~(\ref%
{eq:NSpSqGen}), we only need to prove that $\langle\mathcal{J}_0\rangle$ and 
$(\Delta{}_{\{\theta_i\}})_{\mathrm{min}}^2 $ are invariant. We now use the
well-known fact that, for every unitary transformation $U_i$ on a qubit $i$,
there corresponds a unique $3\times 3$ orthogonal rotation matrix $O_i$ such
that 
\begin{equation}
\vec{\sigma}_i^\prime\equiv U_i^\dagger \vec{\sigma}_i U_i =O_i \vec{\sigma}%
_i, \quad \langle\vec{\sigma}_i^\prime\rangle =O_i \langle\vec{\sigma}%
_i\rangle.  \label{uuu}
\end{equation}
From Eqs. (\ref{angular0}) and (\ref{angular}), we may write the expectation
value $\langle\mathcal{J}_0\rangle$ as 
\begin{equation}
\langle\mathcal{J}_0\rangle=\frac{1}{2}\sum_{i=1}^N | \langle\vec{\sigma}%
_i\rangle|.  \label{J01}
\end{equation}
From Eqs.~(\ref{uuu}) and (\ref{J01}), we observe that the expectation value 
$\langle\mathcal{J}_0\rangle$ is invariant under the local unitary
transformation.

To prove that $(\Delta{}_{\{\theta_i\}})_{\mathrm{min}}^2 $ is invariant, we
first write it as 
\begin{equation}
(\Delta{}_{\{\theta_i\}})_{\mathrm{min}}^2 =\frac{1}{4}\left[%
N+2\,\left(\sum_{i=1}^N\sum_{j>i=1}^N \hat{n}_{\theta_i}^T \mathcal{T}^{(ij)}%
\hat{n}_{\theta_j}\right)_{\mathrm{min}}\right],  \label{newform}
\end{equation}
where 
\begin{equation}
\mathcal{T}^{(ij)}_{\alpha\beta}=
\langle\sigma_{i\alpha}\sigma_{j\beta}\rangle , \quad \quad \alpha,\,
\beta=x,y,z,
\end{equation}
are the matrix elements of the $3\times 3$ correlation matrix~\cite{Hor96}
for a given qubit pair $i,\, j$. Note that $\hat{n}_{\theta_i}$ is a column
vector.

From Eq.~(\ref{uuu}), under the unitary transformations $U$, the parameter $%
\hat{n}_{\theta_i}$ and the correlation matrix $\mathcal{T}^{(ij)}$
transform as follows~\cite{Hor96}: 
\begin{equation}
\hat{n}_{\theta_i}^{\prime}=O_i \hat{n}_{\theta_i},\quad \mathcal{T}%
^{^{\prime}(ij)}=O_i\mathcal{T}^{(ij)}O_j^T.
\end{equation}
Applying the above equation to Eq.~(\ref{newform}) leads to the invariance
of $(\Delta{}_{\{\theta_i\}})_{\mathrm{min}}^2$ under the unitary
transformation $U$. $\Box$

Though the minimization $(\Delta {}_{\{\theta_i\}})_{\mathrm{min}}$ over all
the directions $\hat{n}_{\theta_i}$ (orthogonal to the qubit orientations)
appears to be non-trivial, local invariance of $\tilde{\xi}_{k}$ leads to a
simplified analysis for multi-qubit systems. For instance, we may consider a
product state with each qubit being in a different state. For this case, we
can always find a unitary operator $U$ to transform this state to the
symmetric product state $|0\rangle\otimes\cdots\otimes|0\rangle$, for which
the squeezing parameters $\tilde{\xi}_k$ are easily found to be 1. This
simple example indeed displays that when the quantum correlations do not
exist, the spin squeezing parameter cannot be less than 1. Recognizing that
the both the spin squeezing parameter $\tilde{\xi}_k$ and quantum
entanglement are invariant under local unitary transformations, we next
investigate the relations between spin squeezing and entanglement.

\section{Relations to quantum entanglement}

Spin squeezing is closely related to, and implies, quantum entanglement~\cite%
{Sor01a,Kit01,Dominic2,Sanmore,Xia03}. As the spin squeezing criteria we
propose satisfy local unitary invariance, a close relation between these
criteria and quantum entanglement is expected.

\subsection{Two-qubit states}

To see the relations of the local unitary invariant spin squeezing to
quantum entanglement, let us examine the arbitrary two-qubit pure state
given by 
\begin{equation}  \label{eq:Psi12}
|\Phi_{12}\rangle =\alpha\, |00\rangle +\beta\, |01\rangle +\gamma\,
|10\rangle +\delta\,|11\rangle.
\end{equation}
It has been shown earlier~\cite{Kit01} that for a pure symmetric state of
pair qubits (with $\beta=\gamma$ in Eq.~(\ref{eq:Psi12})), the concurrence
quantifying two-qubit entanglement~\cite{Hill97} is related to the spin
squeezing parameter $\xi_{1}$ of Eq.~(\ref{eq:Kitagawa}) through $\mathcal{C}%
=1-\xi_{1}^2$, thereby implying the equivalence between spin squeezing and
quantum entanglement. However, the squeezing parameter $\xi_1$ cannot apply
to the non-symmetric state $|\Phi_{12}\rangle $ with $\beta \neq \gamma$ as
the parameter can be larger than unity even for a non-symmetric product
state. The spin squeezing criterion ${\xi}_{1}<1$ cannot separate an
entangled state from a non-entangled one. The relationship between the local
invariant spin squeezing parameter $\tilde{\xi}_{k}$ and the concurrence,
for a pure state of two qubits given by Eq.~(\ref{eq:PureStateSpSq2}),
therefore provides a generalization of the earlier result~\cite{Kit01}
applicable to both symmetric as well as non-symmetric pure states of two
qubits.

For the general two-qubit state we have\newline
\textbf{Proposition 2}: \textit{Spin squeezing with local unitary invariance
and quantum entanglement are equivalent for arbitrary two-qubit pure states.
The quantitative relations between spin squeezing parameters and the
concurrence are given by 
\begin{equation}
\tilde{\xi}_{1}=\sqrt{1-\mathcal{C}},\quad \tilde{\xi}_{2}=\frac{1}{\sqrt{%
1+{}}}.  \label{eq:PureStateSpSq2}
\end{equation}%
}\newline
\textit{Proof}: Up to local unitary operations, we can express the state $%
|\Phi _{12}\rangle $ using a Schmidt decomposition through 
\begin{align}
& |\Phi _{12}\rangle =\lambda _{1}|00\rangle +\lambda _{2}|11\rangle , 
\notag  \label{eq:Psi12S} \\
& 0\leq \lambda _{1},\ \lambda _{2}\leq 1,\;\lambda _{1}^{2}+\lambda
_{2}^{2}=1,
\end{align}%
where the Schmidt coefficients $\lambda _{1}$ and $\lambda _{2}$ are related
to $\alpha ,\ \beta ,\ \gamma ,\ \mathrm{and}\ \delta $ by 
\begin{align}
\lambda _{1}^{2}=& \frac{1}{2}\left[ 1+\sqrt{1+4\,|(\beta \gamma -\alpha
\delta )|^{2}}\right] ,  \notag \\
\lambda _{2}^{2}=& \frac{1}{2}[1-\sqrt{1+4\,|(\beta \gamma -\alpha \delta
)|^{2}}].
\end{align}

The mean spin is along the $z$ direction, and it is easy to see that 
\begin{equation}
\langle{}_0\rangle=|\lambda_{1}^2-\lambda_{2}^2|.  \label{aaa}
\end{equation}
Then, from Eq.~(\ref{newform}), we find 
\begin{align}
(\Delta{}_{\{\theta_i\}})^2 =&\frac{1}{2}(1+
\cos\theta_1\cos\theta_2\langle\sigma_{1x}\sigma_{2x}\rangle  \notag \\
&+\sin\theta_1\sin\theta_2\langle\sigma_{1y}\sigma_{2y}\rangle  \notag \\
&+\sin\theta_1\cos\theta_2\langle\sigma_{1y}\sigma_{2x}\rangle  \notag \\
&+\cos\theta_1\sin\theta_2\langle\sigma_{1x}\sigma_{2y}\rangle)  \notag \\
=&\frac{1}2[1+2\cos(\theta_1+\theta_2)\lambda_1\lambda_2],
\end{align}
which manifestly implies that 
\begin{equation}
(\Delta{}_{\{\theta_i\}})_{\mathrm{min}}=\sqrt{\frac{1-2\lambda_1\lambda_2}{2%
}}.  \label{aaa111}
\end{equation}

Therefore, from Eqs.~(\ref{aaa}), (\ref{aaa111}), and the definition of
squeezing parameters $\tilde{\xi}_k$, we have 
\begin{equation}  \label{eq:PureStateSpSq}
\tilde{\xi}_{1}=\sqrt{1-2\lambda_1\lambda_2}, \quad \tilde{\xi}_{2}= \frac{%
\sqrt{1-2\ \lambda_{1}\ \lambda_{2}}} {|(\lambda_{1}^2-\lambda_{2}^2)|}.
\end{equation}
Identifying now that the {concurrence} ${}$~\cite{Hill97} is related to the
Schmidt coefficents $\lambda_{1},\ \lambda_{2}$ through 
\begin{equation}
{}=2\,\lambda_{1}\,\lambda_{2}= 2|(\beta\gamma - \alpha\delta)|,
\end{equation}
and expressing $|(\lambda_{1}^2-\lambda_{2}^2)|=\sqrt{1-{}^2}$ in Eq.~(\ref%
{eq:PureStateSpSq}), we obtain the proposition. $\Box$

This proposition reveals that \emph{a pair of qubits, which share a pure
entangled state (${}\neq 0$), is always in a spin squeezed state}. This
one-to-one relation between spin squeezing and entanglement implies that we
can use $\tilde{\xi}_{k}$ as a measure of entanglement for pure bipartite
states of qubits. However, it should be noted that the maximum spin
squeezing characterized by $\tilde{\xi}_{1}=0$ and $\tilde{\xi}_{2}=1/\sqrt{2%
}$ for ${}=1$ (maximally entangled states) is realised in the limit of ${%
{}\to 1}$, since the qubits have no preferred orientation in space ($\langle%
\vec{\sigma}_i\rangle=0)$ when they share a maximally entangled state.

\subsection{Multi-qubit states}

We now examine the relations between spin squeezing and quantum entanglement
for arbitrary multi-qubit states. For all separable $N$-qubit states, it was
already shown that the original spin squeezing parameter $\xi_2\ge 1$~\cite%
{Sor01a}. However, here the parameter $\tilde{\xi}_2$ is locally unitary
invariant, and therefore more closely related to quantum entanglement as we
have seen in Proposition 2.

For separable states, we have\newline
\textbf{Proposition 3}: \textit{For all separable $N$-qubit states, $\tilde{%
\xi}_2\ge 1$.}\newline
\textit{Proof}: Along similar lines of proof $\xi_2\ge 1$~\cite{Sor01a}, we
now give the proof. A separable state can be written as 
\begin{equation}  \label{eq:SepNonSym}
\rho=\sum_{k}p_{k}\, \rho_1^{(k)}\otimes\rho_2^{(k)}\otimes \ldots
\otimes\rho_N^{(k)},\ \ \sum_{k}p_k=1.
\end{equation}
Calculating the variance of ${}_{\{\theta_i\}}$, we find 
\begin{align}  \label{eq:SepSpSq1}
(\Delta {}_{\{\theta_i\}})^2&=\frac{N}{4}+ \frac{1}{4}\sum_{k}
p_{k}\sum_{i=1}^{N}\sum_{j\neq i=1}^{N}\langle \vec{\sigma}_i\cdot \hat{n}%
_{\theta_i}\rangle_{k}\langle\vec{\sigma}_j \cdot \hat{n}_{\theta_j}%
\rangle_{k}  \notag \\
&=\frac{N}{4}+\frac{1}{4}\sum_{k}p_{k}\sum_i\sum_j \langle \vec{\sigma}%
_i\cdot \hat{n}_{\theta_i}\rangle_{k}\langle \vec{\sigma}_j\cdot \hat{n}%
_{\theta_j}\rangle_{k}  \notag \\
& \ \ \ \ \ \ -\frac{1}{4} \sum_{k} p_{k}\sum_i\langle \vec{\sigma}_i\cdot 
\hat{n}_{\theta_i}\rangle^2_{k}  \notag \\
& = \frac{N}{4}+\frac{1}{4}\sum_{k}p_{k}\left(\sum_{i}\langle \vec{\sigma}%
_{i}\cdot\hat{n}_{\theta_i}\rangle_{k}\right)^2  \notag \\
&\ \ \ \ \ \ -\frac{1}{4} \sum_{k} p_{k}\sum_i\langle \vec{\sigma}_i\cdot 
\hat{n}_{\theta_i}\rangle ^2_{k}  \notag \\
& \geq \frac{N}{4}-\frac{1}{4}\sum_{k}p_{k}\sum_{i}\langle\vec{\sigma}%
_{i}\cdot\hat{n}_{\theta_i}\rangle _k ^2.
\end{align}
Using the condition 
\begin{equation}
\langle \vec{\sigma}_i \cdot\hat{n}_{\theta_i}\rangle_k^2 +\langle\vec{\sigma%
}_i\cdot \hat{n}^{\prime}_{\theta_i}\rangle_k^2+\langle\vec{\sigma}_i\cdot%
\hat{n}_{i0} \rangle_k^2\leq 1
\end{equation}
with 
\begin{equation}
\hat{n}^{\prime}_{\theta_i}=-\sin(\theta_i)\hat{n}_{i\perp}+\cos(\theta_i)%
\hat{n}_{i\vdash},
\end{equation}
we obtain 
\begin{align}  \label{SepSpSq2}
&\frac{N}{4}-\frac{1}{4}\sum_{k}p_{k}\sum_{i}\langle\vec{\sigma}_{i}\cdot%
\hat{n}_{\theta_i}\rangle _k ^2  \notag \\
\geq &\frac{1}{4}\sum_{k} p_{k}\sum_i (\langle \vec{\sigma}_i \cdot \hat{n}%
^{\prime}_{\theta_i}\rangle_{k}^2 + \langle \vec{\sigma}_i\cdot \hat{n}%
_{i0}\rangle_{k}^2)  \notag \\
\geq & \frac{1}{4}\sum_{k} p_{k} \sum_i\langle \vec{\sigma}_i \cdot \hat{n}%
_{i0}\rangle_{k}^2  \notag \\
=&\frac{1}{N}\sum_{k,i}\sqrt{p_k}^2 \sum_{k,i} \left(\sqrt{p_k}\frac{\langle 
\vec{\sigma}_i \cdot \hat{n}_{i0}\rangle_{k}^2}{2}\right)^2  \notag \\
\ge & \frac{1}{4N}\left(\sum_{k} p_{k} \sum_i\langle \vec{\sigma}_i\cdot 
\hat{n}_{i0}\rangle_{k}\right)^2  \notag \\
=&\frac{\langle{}_0\rangle^2}{N},
\end{align}
where the last inequality follows from the Cauchy-Schwarz inequality, and
the last equality from 
\begin{equation}
\langle{}_0\rangle^2= \displaystyle\frac{1}{4}\left(\sum_{k} p_{k}
\sum_i\langle \vec{\sigma}_i\cdot \hat{n}_{i0}\rangle_{k}\right)^2
\end{equation}
in the separable state.

Therefore, we obtain the inequality 
\begin{equation}  \label{eq:SepSpSq5}
(\Delta {}_{\{\theta_i\}})^2 \geq \frac{1}{N}\, \langle {}_0\rangle ^{2},
\end{equation}
which leads to the result that $\tilde{\xi}_{2}\geq 1$ for all separable
states. $\Box$

In this way, local unitary invariant spin squeezing characterized by $\tilde{%
\xi}_2<1$ serves as a sufficient condition for quantum entanglement.

\section{Conclusion}

In conclusion, we have proposed a local invariant criterion of spin
squeezing in multi-qubit systems, which is applicable for arbitrary
multi-qubit states. Due to the local unitary invariant property of the spin
squeezing criteria and quantum entanglement, we find that (i)~for arbitrary
two-qubit states, spin squeezing is equivalent to quantum entanglement, and
(ii)~for arbitrary multi-qubit states, we find that the squeezing parameter $%
\tilde{\xi}_2<1$ implies quantum entanglement. The inequality $\tilde{\xi}%
_2<1$ serves as a sufficient condition for entanglement.

\begin{acknowledgments}
ARU acknowledges the support of a Macquarie University Visiting Fellowship.
We acknowledge valuable comments from Prof.\ Li You and helpful discussions
with Dominic W. Berry. This project has been supported by an Australian
Research Council Large Grant and by an Australian Department of Education,
Science and Training Innovation Access Program Grant to support the European
Fifth Framework Project QUPRODIS.
\end{acknowledgments}

\end{document}